\documentclass[12pt,a4paper]{article}
\usepackage[latin1]{inputenc}
\usepackage{amsmath}
\usepackage{amsfonts}
\usepackage{amssymb}
\usepackage{hyperref}
\hypersetup{colorlinks=true}
\usepackage{cite}
\begin{document}
\author{Levent Akant${^\dag}$, Emine Ertu\u{g}rul${^\ddag}$,\\ Ferzan Tapramaz${^\S}$, O. Teoman Turgut${^*}$ \\ Department of Physics, Bo\u{g}azi\c{c}i University \\ 34342 Bebek, Istanbul, Turkey \\ $^\dag$levent.akant@boun.edu.tr, $^\ddag$emine.ertugrul@boun.edu.tr,
\\${^\S}$waskhez@gmail.com,${^*}$turgutte@boun.edu.tr}
\title{\bf Bose-Einstein Condensation on a Manifold with Nonnegative Ricci Curvature}
\maketitle
\begin{abstract}
The Bose-Einstein condensation for an ideal Bose gas and for a
dilute weakly interacting Bose gas in a manifold with nonnegative Ricci
curvature is investigated using the heat kernel and eigenvalue
estimates of the Laplace operator. The main focus is on the nonrelativistic gas. However, special relativistic ideal gas is also discussed. The thermodynamic limit of the heat kernel and eigenvalue estimates is taken and the results are used to derive
bounds for the depletion coefficient. In the case of a weakly interacting gas Bogoliubov approximation is employed. The ground state is analyzed using heat kernel methods and finite size effects on the ground state energy are proposed. The justification of the c-number substitution on a manifold is given.
\end{abstract}

\newpage

\section{Introduction}
The purpose of this paper is to analyze the depletion coefficient
(number density of bosons out of the condensate) and the ground state energy of the
Bose-Einstein condensate on a Riemannian manifold with a nonnegative
Ricci curvature using global heat kernel and eigenvalue estimates and also the heat kernel asymptotics
for the Laplacian on the manifold. The basic observation we make is that both the depletion coefficient and the ground state energy can be expressed
in terms of the heat kernel of the Laplacian. Then we exploit this fact and analyze the depletion coefficient and the ground state energy using the
above mentioned bounds and asymptotics.  Applying our results to the flat space case we derive finite size corrections to the ground state energy of the weakly
interacting Bose gas.

We will consider both the ideal gas and the gas with a weak
hard-core repulsion \cite{Bogoliubov}, \cite{HuangYang},
\cite{LeeYang}, \cite {LeeHuangYang}. The latter case will be
analyzed using the curved space version of the Bogoliubov theory. The Bogoliubov approximation scheme starts with the replacement of the
ground state creation and annihilation operators by c-numbers. One then truncates the Hamiltonian, diagonalizes the resulting quadratic Hamiltonian by a Bogoliubov transformation
and thus derives the Bogoliubov spectrum. That
the first step of this procedure, the c-number substitution, is an exact procedure and not merely an approximation was
proven by several authors \cite{Ginibre}, \cite{Lieb}, \cite{Suto1},
\cite{Suto2}. In particular, in \cite{Lieb} Lieb et al. have shown
that the c-number substitution is a general property of second
quantization. The generalization of this property to the Riemannian
manifolds will be discussed in the Appendix B. The use of the
Neumann boundary conditions for the Bose gas implies that the one
particle ground state wave function is constant. Since the ideal gas
is noninteracting, the thermodynamic properties are sensitive to the
boundary conditions imposed, in a sense all the interaction comes
from the boundary. The use of Neumann boundary condition is equally
acceptable as the Dirichlet one, the former would mean that the
particle flux from the boundary is zero. In our approach the Neumann
boundary condition is more natural; in fact, in the case of the ideal gas all the results
that we derive using the Neumann boundary conditions can also
be obtained for Dirichlet boundary conditions. The homogeneity of the ground
state allows us to carry the flat space proof to the curved case.
For an extensive review of the Bogoliubov theory in flat space see
\cite{Zagrebnov} and references therein (for a curved space version
 see \cite{Rivas}). For a mathematically rigorous account of
Bose-Einstein condensation in flat space see \cite{Liebbook}, for a more
traditional approach see the monographs \cite{Pit}, \cite{Pethick}, and for finite temperature gases on curved spaces see
\cite{Dowker1, Dowker2, T1, T2, T3}.

The effect of trapping potentials will be discussed elsewhere. We
will mainly consider the non-relativistic gas. However, in the case
of an ideal gas we will discuss the special relativistic
generalization, \cite{Jutner, Glaser, Landsberg, Nieto, Altaie,
Beckmann, Carvalho, Beckmann2, Haber, Haber2, Burakovsky, Filippi}
as well.

For an ideal gas in a flat manifold the behavior of the chemical
potential $\mu$ is crucial for the understanding of the
Bose-Einstein condensation. The chemical potential is determined by
requiring the constancy of the total number density of the Bose gas.
Below a critical temperature $T_{c}$ the chemical potential is of
the order of $V^{-1}$ where $V$ is the volume of the gas \footnote{As the chemical potential is an intensive variable its dependence of $N$ and $V$ is of the form
$\mu=f(N/V,T)+O(V^{-\lambda})$ where $\lambda$ is a positive coefficient. However at $T=T_{c}$ the leading term vanishes and this gives rise to a volume dependence of the chemical potential.}. This is a
direct consequence of the fact  that for $T\leq T_{c}$
\begin{eqnarray}
    n_{0}=\lim_{V\rightarrow
    \infty}\frac{1}{V}\frac{1}{e^{-\beta\mu}-1}= \lim_{V\rightarrow
    \infty}\frac{1}{V\beta|\mu|}>0.
\end{eqnarray}
Here we assumed that the ground state energy is zero. In the flat space, when we consider a box of typical dimensions given by $L=O(V^{1/3})$, we have
\begin{eqnarray}\label{asymp}
    \epsilon_{\sigma}=O(L^{-2})\;\;\textrm{for}\;\;\sigma\neq
    0.
\end{eqnarray}
Here $\sigma$'s are the quantum numbers of the gas, with $\sigma=0$
denoting the ground state, and $\epsilon_{\sigma}$'s are the
corresponding energies. Therefore, we see that
for $T<T_c$, for $\sigma\neq 0$
\begin{eqnarray}
    n_{\sigma}\;\leq \;\lim_{V\rightarrow
    \infty}\,\frac{1}{V}\,\frac{1}{e^{\beta \epsilon_{\sigma}}-1}=0.
\end{eqnarray}
So in the thermodynamic limit the ground state is occupied
macroscopically, while the excited states are not. As we will see,
thanks to eigenvalue bounds on manifolds, this argument remains
correct for {\it macroscopically admissible volumes}, which we define to
be a domain inside the manifold, such that the diameter $D_M$ is
also $O(V^{1/d})$, where $d$ is the dimension of the manifold. More
precisely,  on a Riemannian manifold $S$ with a non-negative Ricci
curvature the eigenvalues of the Laplacian obey precisely
(\ref{asymp}) \cite{LiYau}, \cite{Colbois} if $L$ is interpreted as
the geodesic diameter $D_M$ of the confining box which is abstracted
as a submanifold $M\subset S$ with boundary. We will assume that the
gas obeys Neumann boundary conditions on $\partial M$.

Our main observation is that the number density of the excited
particles can be expressed in terms of the trace of the heat kernel
on $S$. To see that this is so is not difficult in the case of an
ideal gas, but is more involved in the case of a weakly interacting
gas. Once we establish this result we will use the heat kernel and
eigenvalue bounds of the Laplacian to discuss the Bose-Einstein
condensation on $S$. In this paper we will not attempt any
comparison with the existing bounds on the depletion coefficient in
flat space \cite{Hohenberg, Roepstorff, Penrose, Pitaevskii1,
Pitaevskii2, Leggett}. Our bounds, which are very geometric in
character, seem to provide a new class of constraints on the
depletion of the condensate. Finally we note the following
alternative approaches to the Bose-Einstein condensation in curved
spaces \cite{Kirsten1},\cite{Kirsten2}.

\section{Heat Kernel and Eigenvalue Bounds}

We start by introducing our geometric setting and by summarizing the
existing heat kernel and eigenvalue estimates that will play
important roles in our subsequent discussions. For the sake of
completeness and in order to clarify certain relations between the
estimates, we will
present more detailed discussion of some of these results in
Appendix A, following the mathematics literature.

Let $(S,g)$ be a $d$ dimensional Riemannian manifold with metric $g$
and nonnegative Ricci curvature \footnote{A symmetric covariant
tensor field $T$ of rank 2 is bounded below if there exist a real
number $c$ such that for any vector field $X$ on $S$
\begin{equation}
    T(X,X)\geq c \,g(X,X).
\end{equation}
In this case one writes $T\geq c$.}
\begin{equation}
    \textbf{Ric}_{S}\geq 0.
\end{equation}
Let $M$ be a connected open submanifold of $S$ with compact closure
and smooth convex boundary $\partial M$. Clearly the Ricci curvature
of $M$ is also nonnegative. Let $\Delta$ be the Laplacian of the
metric $g$ and $\{f_{\sigma}\}$ ($\sigma=0,1,2,\ldots$) be a
complete orthonormal set of real (standing wave) square-integrable
eigenfunctions of $-\Delta$ on $M$, obeying the Neumann boundary
conditions
\begin{equation}
    -\Delta\,f_{\sigma}=\epsilon_{\sigma}f_{\sigma},\;\;\;\;\left.\hat{n}\cdot \nabla f_{\sigma}\right|_{\partial
    K}=0.
\end{equation}
Here $\hat{n}$ is the outward looking unit normal to $\partial M$.
The eigenvalues can be ordered as
\begin{equation}
    \epsilon_{0}< \epsilon_{1}\leq \ldots \rightarrow\infty.
\end{equation}
The ground state is
\begin{equation}
f_{0}=\frac{1}{\sqrt{V}},
\end{equation}
with eigenvalue $\nu_{0}=0$. Here $V$ is the volume of $M$.
Connectedness of $M$ implies the uniqueness of the ground state and
the existence of the fundamental gap $\epsilon_{1}>0$.

Note that the reality of the eigenfunctions implies
\begin{equation}\label{ort1}
    \int d\mu_{g}(x) f^{*}_{\sigma}(x)f_{\rho}(x)=\int d\mu_{g}
    f_{\sigma}(x)f_{\rho}(x)=\delta_{\sigma\rho},
\end{equation}
and for $\sigma\neq 0$ we also have
\begin{equation}\label{ort2}
    \int d\mu_{g}(x) f_{\sigma}(x)=\sqrt{V}\int d\mu_{g}
    f_{0}(x)f_{\sigma}(x)=0.
\end{equation}
Here $d\mu_{g}$ is the Riemannian volume element corresponding to
the metric $g$.

For the Neumann heat kernel on a manifold $M$ with a nonnegative
Ricci curvature and diameter $D_{M}$\footnote{ The diameter $D_{M}$
of a Riemannian manifold $M$ is given by $D_{M}=\sup\{r(p,q):p,q\in
M\}$ where $r$ is the geodesic distance on $M$. } one has the
following estimates of Li and Yau \cite{LiYau}
\begin{equation}\label{bound1}
  \frac{1}{(4\pi t)^{d/2}}V \leq \textrm{Tr}\,{e^{\Delta t}}\leq
  \tilde{C}(d)g(t).
\end{equation}
Here $\tilde{C}(d)$ is a positive constant which depends only on the
dimension $d$ and
\[g(t)=\left \{  \begin{array}{ll}
   \left(\frac{D_{M}}{\sqrt{t}}\right)^{d} & \mbox{if $\sqrt{t}\leq D_{M}$},  \\
1 & \mbox{if $\sqrt{t}\geq D_{M}$}. \end{array}\right.\]

A direct consequence of the upper bound for the trace of the heat
kernel is the eigenvalue bound \cite{LiYau}
\begin{equation}\label{liyau}
    \epsilon_{\sigma}\geq \frac{C(d)}{D_{M}^{2}}\,(\sigma+1)^{2/d}\geq
    \frac{C(d)}{D_{M}^{2}}\,\sigma^{2/d},
\end{equation}
where $C(d)$ is a positive constant which depends only on the
dimension.

On the other hand one also has the following upper bound of Colbois
and Maerten \cite{Colbois} for the eigenvalues
\begin{equation}\label{colbois}
    \epsilon_{\sigma}\leq B(d) \left(\frac{\sigma}{V}\right)^{2/d}.
\end{equation}
 Here $B(d)$ is a positive constant which depends only on the
 dimension.

Using the eigenvalue bounds
(\ref{liyau}) and (\ref{colbois}) we get the following bounds for $\textrm{Tr}^{\prime}e^{\Delta t}$ (here prime means the ground state is omitted in the trace)
\begin{equation}
   \frac{1}{V} \sum_{\sigma=1}^{\infty}e^{-\frac{t B}{V^{2/d}}\,\sigma^{2/d}}\leq \frac{1}{V}\sum_{\sigma=1}^{\infty}e^{-t\epsilon_{\sigma}}\leq
    \frac{1}{V}\sum_{\sigma=1}^{\infty}e^{-\frac{t C}{D_{M}^{2}}\,\sigma^{2/d}}.
\end{equation}
Bounding the series by integrals we get
\begin{equation}
    \int_{1}^{\infty}dx\,e^{-t B x^{2/d}}\leq
    \frac{1}{V}\,\textrm{Tr}'e^{\Delta t}\leq \frac{D_{M}^{d}}{V}\int_{0}^{\infty}dx\,e^{-t C x^{2/d}}.
\end{equation}
The integrals can be evaluated explicitly
\begin{equation}\label{important}
    \frac{1}{B^{d/2}}\,\frac{d}{2}\,\Gamma\left(\frac{d}{2},\frac{t\,B}{V^{2/d}}\right)\left(\frac{1}{t
    }\right)^{d/2}\leq \frac{1}{V}\,\textrm{Tr}'e^{\Delta t}\leq
    \frac{D_{M}^{d}}{V\,C^{d/2}}\,\Gamma\left(\frac{d}{2}+1\right)\left(\frac{1}{t}\right)^{d/2}.
\end{equation}
 Here $\Gamma(x,y)$ is the incomplete gamma function.

Another upper bound for the trace of the heat kernel which holds
for large $t$ is \cite{Wang}
\begin{equation}\label{in}
    \textrm{Tr}^{\prime}e^{\Delta\,t} \leq
    (\textrm{Tr}^{\prime}e^{\Delta\,t_{0}})e^{-\epsilon_{1}(t-t_{0})}\;\;\;t\geq t_{0}.
\end{equation}
Here $t_{0}$ is some fixed time.

\section{Ideal Bose Gas in a Riemannian Manifold}

The single particle free Hamiltonian is taken as
\begin{equation}
    h=-\Delta.
\end{equation}
The corresponding many-body Hamiltonian is
\begin{equation}
    H_{0}=\int d\mu_{g}
    \phi^{\dag}(x)h\,\phi(x).
\end{equation}
The thermal averages in the grand-canonical ensemble are given by
\begin{equation}
\langle
O\rangle=\frac{\textrm{Tr}\,O\,e^{-\beta\,H}}{\textrm{Tr}\,e^{-\beta\,H}}.
\end{equation}
Here
\begin{equation}
    H=\int d\mu_{g}
    \phi^{\dag}(x)(h-\mu)\,\phi(x)
\end{equation}
and $\mu$ is the chemical potential.

Expanding $H$ in terms of creation-annihilation operators and normal
ordering the result we get
\begin{equation}
    H=\sum_{\sigma}(\epsilon_{\sigma}-\mu)a_{\sigma}^{\dag}a_{\sigma}.
\end{equation}
Thus the mean occupation numbers are given exactly as in the flat
case
\begin{equation}
    N_{\sigma}=\langle
    a_{\sigma}^{\dag}a_{\sigma}\rangle=\frac{1}{e^{\beta(\epsilon_{\sigma}-\mu)}-1},
\end{equation}
where we have taken, as usual, $\mu<0$. Clearly, $N_{0}$ is singular
at $\mu=0$. As in the flat case, fixing the total number of particles allows us to determine $\mu$.
To maximize the occupation of each level, we should let $\mu\to0^-$. However,  that is a delicate limit, the chemical potential is not strictly zero,  the macroscopic occupation of the ground state again leads to
$\mu=O(V^{-1})$ so that we find a well-defined thermodynamic limit.
Note that
\begin{equation}
N=\sum_\sigma  \frac{1}{e^{\beta(\epsilon_{\sigma}-\mu)}-1},
\end{equation}
does not scale in a simple way to determine the chemical potential in a simple way even in the continuum limit, since the density of states may be a complicated expression in general.

\textit{Finite Volume:} First note that
\begin{equation}
 n_{\sigma}=\frac{N_{\sigma}}{V} =\frac{1}{V}\frac{e^{-\beta(\epsilon_{\sigma}-\mu)}}{1-e^{-\beta(\epsilon_{\sigma}-\mu)}}=\frac{1}{V}
  \sum_{k=1}^{\infty}e^{k\beta\mu}e^{-k\beta\epsilon_{\sigma}},
\end{equation}
and consequently,
\begin{eqnarray}\label{ne}
    n_{e}(T):=\sum_{\sigma\neq 0}n_{\sigma}&=& \frac{1}{V}\sum_{\sigma\neq
    0}\sum_{k=1}^{\infty}e^{k\beta\mu}e^{-k\beta\epsilon_{\sigma}}\nonumber\\
    &=&\frac{1}{V}\sum_{k=1}^{\infty}e^{k\beta\mu}\,\textrm{Tr}'e^{-k\beta\,h}.
\end{eqnarray}
Here $\textrm{Tr}'$ denotes the trace with the ground state omitted.
We employ (\ref{in}) to study the low temperature behavior of
$N_{e}(T)$. Fixing $\beta_{0}$ such that $\beta>\beta_{0}$ we see
that
\begin{eqnarray}\label{numberbound}
 n_{e}(T) \leq
 \frac{1}{V}\,e^{\epsilon_{1}\beta_{0}}\,(\textrm{Tr}'e^{-h\beta_{0}})\,\sum_{k=1}^{\infty}e^{-k\epsilon_{1}\beta}=
 \frac{1}{V}(\textrm{Tr}'e^{-h\beta_{0}})\,
 \frac{e^{\epsilon_{1}\beta_{0}}}{e^{\epsilon_{1}\beta}-1}.
\end{eqnarray}
So by choosing $T$ low enough we can make $n_{e}$ less than any
preassigned value of $n$. Then the particles in excess, whose density is $n-n_{e}$,
form condensate in the ground state.

The above analysis is independent of the dimension
of the manifold $M$ and the bound on the Ricci curvature. However,
the right hand side of (\ref{numberbound}) is in general divergent
as $V\rightarrow\infty$. So the bound is useless in the
thermodynamic limit.

\textit{Thermodynamic Limit in Terms of the Heat Kernel:} Using the upper bound given in (\ref{bound1}) we see that
\begin{eqnarray}\label{nbound}
    n_{e}(T)&\leq & \frac{\tilde{C}(d)}{V}\sum_{k=1}^{\infty} e^{-k\beta |\mu|}g(k\beta)\\
&\leq &
\frac{\tilde{C}(d)\,D_{M}^{d}}{V}\frac{1}{\beta^{d/2}}\sum_{k=1}^{[D_{M}^{2}/\beta]}
\frac{e^{-k\beta |\mu|}}{k^{d/2}}+
\frac{\tilde{C}(d)}{V}\sum_{k=[D_{M}^{2}/\beta]+1}^{\infty}
e^{-k\beta |\mu|}.
\end{eqnarray}
Here square brackets mean integer part. Now the second term is just a geometric series whose sum is
\begin{equation}
    \frac{\tilde{C}(d)}{V}\,\frac{e^{-\left(\left[\frac{D_{M}^{2}}{\beta}\right]+1\right)\beta |\mu|}}{1-e^{-\beta |\mu|}}.
\end{equation}
 Using $\mu=O(V^{-1})$ we see that this goes to zero in the thermodynamic limit $V, D_{M}\rightarrow\infty$.

Our basic assumption regarding the thermodynamic limit will be the
following asymptotic relation between the volume and the diameter of
our box $M$
\begin{equation}
D_{M}=O(V^{1/d})\;\;\;\textrm{as}\;\;\;V\rightarrow\infty.
\end{equation}
This is a nontrivial condition and it is not so obvious if we can satisfy this on a Ricci nonnegative manifold. If we have a strictly positive lower bound for the Ricci,  by Myers' theorem the manifold necessarily becomes compact. In our case, due to the Bishop-Gromov volume coomparison theorem, the geodesic balls in a Ricci non-negative space  cannot have volumes growing faster than the flat case \cite{Gallot}. This is important since, a natural  set of   boxes to consider would be geodesic balls. It is known
that on a complete Riemannian manifold ${\cal M}$ of dimension $n$,  with nonnegative Ricci , there is a constant $\epsilon(n)$  such that if for some point $p$ we have
\begin{equation}
{\rm Vol} B_r(p)\geq (1-\epsilon(n)) \Omega_n r^n
\end{equation}
for all $r$, where $\Omega_n$ refers to the volume of the standart unit $n$-ball, then ${\cal M}$ is diffeomorphic to the Euclidean space, but not necessarily isometric to it. So clearly there are some interesting examples within our class of manifolds. For further results and references related to this subject  we refer to the review article \cite{wei}.

Under this assumption, in the thermodynamic limit we get
\begin{equation}
     n_{e}(T)\leq \tilde{C}(d)\frac{A}{\beta^{d/2}} \sum_{k=1}^{\infty}\frac{e^{-k\beta |\mu|}}{k^{d/2}}\leq
     \tilde{C}(d)\frac{A}{\beta^{d/2}} \sum_{k=1}^{\infty}\frac{1}{k^{d/2}}=\tilde{C}(d)\frac{A}{\beta^{d/2}}\,\zeta\left(\frac{d}{2}\right).
\end{equation}
Here
\begin{equation}
    A=\lim_{V\rightarrow\infty}\frac{D_{M}^{d}}{V}.
\end{equation}
For $d\geq 3$ $n_{e}(T)$ is clearly finite and vanishes as
$T\rightarrow 0$.

For $d=2$ the above bound is of no use in the thermodynamic limit. In order to deduce the behavior of $n_{e}(T)$ at $d=2$ we use the lower bound
given in (\ref{bound1}).
\begin{eqnarray}
  n(T) &=& \frac{1}{V}\sum_{k=1}^{\infty} e^{-k\beta |\mu|}\left(\textrm{Tr}e^{-k\beta\,h}\right) \\
   &=& \frac{1}{V}\sum_{k=1}^{\infty} e^{-k\beta |\mu|}\,\textrm{Tr}e^{-k\beta\,h} \\
   &\geq & \frac{1}{(4\pi \beta)^{d/2}}\sum_{k=1}^{\infty}\frac{ e^{-k\beta |\mu|}}{k^{d/2}}.
\end{eqnarray}
By a simple integral test the series is seen to be larger than
\begin{equation}
    \int_{1}^{\infty}dx\,\frac{e^{-x\beta |\mu|}}{x^{d/2}}=
    (\beta|\mu|)^{d/2-1}\int_{\beta|\mu|}^{\infty}
    dy\,\frac{e^{-y}}{y^{d/2}}=(\beta|\mu|)^{d/2-1}\,\Gamma(1-d/2,\beta|\mu|).
\end{equation}
Here $\Gamma(x,y)$ is the incomplete gamma function. If it was the
case that $\mu=O(V^{-1})$ then for $d=2$ we would have a divergent
$n$, which is a contradiction. Thus $\mu\neq O(V^{-1})$ and
condensation does not take place in two dimensions.

\textit{Thermodynamic Limit in Terms of Eigenvalues:} Now using the bounds (\ref{liyau}) and (\ref{colbois}) on the eigenvalues of the Laplacian we get
\begin{equation}
    \sum_{\sigma=1}^{\infty}\frac{1}{V}\frac{1}{e^{\beta\left(\frac{B}{V^{2/d}}\,\sigma^{2/d}-\mu\right)}-1}\leq n_{e}\leq
    \sum_{\sigma=1}^{\infty}\frac{1}{V}\frac{1}{e^{\beta\left(\frac{C}{D^{2}}\,\sigma^{2/d}-\mu\right)}-1}.
\end{equation}

But $\mu=O(V^{-1})$ and the depletion coefficient
can be bound in the $V, D_{M} \rightarrow \infty$ limit as
\begin{equation}
    \int_{0}^{\infty}\,dx
    \frac{1}{e^{\beta\,B\,x^{2/d}}-1}\leq n_{e}\leq A\int_{0}^{\infty}\,dx
    \frac{1}{e^{\beta\,C\,x^{2/d}}-1},
\end{equation}
or after a change of variable
\begin{equation}
    \frac{d}{2(\beta B)^{d/2}}\int_{0}^{\infty}\,d\varepsilon\,
    \frac{\varepsilon^{\frac{d}{2}-1}}{e^{\varepsilon}-1}\leq n_{e}\leq \frac{Ad}{2(\beta\, C)^{d/2}}\int_{0}^{\infty}\,d\varepsilon\,
    \frac{\varepsilon^{\frac{d}{2}-1}}{e^{\varepsilon}-1}.
\end{equation}
Note that the numerator of the integrand is in accordance with the Weyl asymptotic formula for the eigenvalue density of the Laplacian (see e.g. \cite{Chavel}). Since
\begin{equation}
    \int_{0}^{\infty}\,d\varepsilon\,
    \frac{\varepsilon^{\frac{d}{2}-1}}{e^{\varepsilon}-1}=\Gamma\left(\frac{d}{2}\right)\zeta\left(\frac{d}{2}\right),
\end{equation}
we get
\begin{equation}
    \frac{d}{2(\beta B)^{d/2}}\,\Gamma\left(\frac{d}{2}\right)\zeta\left(\frac{d}{2}\right)\leq n_{e}\leq \frac{Ad}{2(\beta\, C)^{d/2}}\,\Gamma\left(\frac{d}{2}\right)\zeta\left(\frac{d}{2}\right).
\end{equation}
On the other hand in flat space
\begin{equation}
    n_{e}^{flat}=S(d)\frac{1}{\beta^{d/2}}\Gamma\left(\frac{d}{2}\right)\zeta\left(\frac{d}{2}\right).
\end{equation}
Here $S(d)$ is the usual density of states factor in
$\mathbf{R}^{d}$
\begin{equation}
    S(d)=V(\mathbf{S}^{d-1})\frac{m^{d/2}}{2^{d/2+1}\pi^{d}}.
\end{equation}
So
\begin{equation}
    \left(\frac{d}{2S(d) B^{d/2}}\right)\,n_{e}^{flat}\leq n_{e}\leq\left( \frac{Ad}{2 S(d) C^{d/2}}\right)\, n_{e}^{flat}.
\end{equation}
Thus we see that $n_{e}$ is divergent for $d\leq 2$ and convergent
for $d>2$. Moreover for $d> 2$, $n_{e}\rightarrow 0$ as
$T\rightarrow 0$ and we have Bose-Einstein condensation at low
temperatures. In fact the critical temperature can be bound as
\begin{equation}
  \left[\frac{Ad}{2nC^{d/2}}\,\Gamma\left(\frac{d}{2}\right)\zeta\left(\frac{d}{2}\right)\right]^{-2/d}
    \leq k_{B}T_{c}\leq
    \left[\frac{d}{2nB^{d/2}}\,\Gamma\left(\frac{d}{2}\right)\zeta\left(\frac{d}{2}\right)\right]^{-2/d}.
\end{equation}

\section{Relativistic Ideal Gas}

In this section we will discuss briefly the use of the heat kernel
method in the case of a relativistic ideal Bose gas. Consider a $1+3$
dimensional manifold with an ultra-static metric
\begin{equation}
    ds^{2}=-dt^{2}+h_{ij}dx^{i}dx^{j}.
\end{equation}
Here $\partial_{t}h_{ij}=0$. The Ricci curvature tensor $^{3}R_{ij}$
of the space-like slices is assumed to be nonnegative. The
equilibrium number density of the excited particles is given by (see e.g. \cite{Hakim})
\begin{eqnarray}
    n_e(T)=\frac{1}{V}\sum\limits_{\sigma\neq 0}\left[\frac{1}{e^{(\beta^{2} \lambda_{\sigma}+\beta^{2}m^{2})^{1/2}-\beta\mu}-1}-
    \frac{1}{e^{(\beta^{2}
    \lambda_{\sigma}+\beta^{2}m^{2})^{1/2}+\beta\mu}-1}\right].
\end{eqnarray}
Here $\lambda_{\sigma}$'s are the eigenvalues of the Neumann problem
for $-^{(3)}\Delta$ on the space-like slices of our metric. Using the
subordination identity
\begin{eqnarray}
  e^{-b\sqrt{x}} &=&
  \frac{b}{2\sqrt{\pi}}\int_{0}^{\infty}\frac{ds}{s^{3/2}}\,e^{-\frac{b^{2}}{4s}}\,e^{-sx},
\end{eqnarray}
we see that
\begin{eqnarray}
n_e (T)= \sum\limits_{k=-\infty}^\infty \int_0^\infty\
\frac{ds}{s^{3/2}}\,\frac{k}{2\sqrt{\pi}}\, e^
{\frac{-k^{2}}{4s}+k\beta\mu}\,\frac{1}{V}(\textrm{Tr}' e^{s\beta^{2}\Delta})
e^{-s\beta^{2}m^{2}},
\end{eqnarray}
which could also be written as
\begin{equation}
n_e (T)= \sum\limits_{k=1}^\infty \int_0^\infty
\frac{ds}{s^{3/2}}\,\frac{k}{\sqrt{\pi}}\, e^
{-\frac{k^{2}}{4s}} \sinh(k\beta\mu)\,\frac{1}{V}(\textrm{Tr}' e^{s\beta^{2}\Delta})
e^{-s\beta^{2}m^{2}}.
\end{equation}
Now using (\ref{important}) in the thermodynamic limit and noticing
\begin{equation}
    \int_{0}^{\infty}\frac{ds}{s^{3}}\,e^{-\frac{k^{2}}{4s}-s\beta^{2}m^{2}}=\frac{8 \beta^{2}m^{2}}{k^{2}}K_{2}(k\beta m),
\end{equation}
we get
\begin{equation}
 \frac{3m^{2}}{B^{3/2}\beta}\sum_{k\neq 0}\frac{e^{k\beta \mu}}{k}K_{2}(k\beta m)   \leq n_{e}(T)\leq \frac{3Am^{2}}{C^{3/2}\beta}\sum_{k\neq 0}\frac{e^{k\beta \mu}}{k}K_{2}(k\beta m).
\end{equation}
Here $K_{2}$ is the modified Bessel function of the second kind.
We recall that $K_\nu$ has the following integral representation for $\nu>0$ and $x>0$,
\begin{equation}
K_\nu(x)=\Big( {\pi\over 2x}\Big)^{1/2} {e^{-x}\over \Gamma(\nu+1/2)}\int_0^\infty ds e^{-s}s^{\nu-1/2}\Big(1+{s\over 2x}\Big)^{\nu-1/2}.
\end{equation}
Using simple estimates we get an upper bound,
\begin{equation}
K_2(x)< {4\over 3\sqrt{2x}}e^{-x} \left[\Gamma(5/2)+{\Gamma(7/2)\over x} +{\Gamma(9/2)\over 4x^2}\right].
\end{equation}
Moreover, by restricting the above sum to only the positive values of $k$,  we get an upper bound. Note that the terms of this series are all decreasing,
and we also emphasize that $\mu<m$ and the maximum is achieved as $\mu\to m^-$.
\begin{eqnarray}
n_e(T)&<&\frac{3Am^{2}}{C^{3/2}\beta}e^{\beta \mu}K_{2}(\beta m) \nonumber \\
&\ & +{4A m^2\over C^{3/2}\beta}\int_1^\infty dx {e^{-x\beta (m-\mu)}\over (\beta m)^{1/2}x^{3/2}} \left[\Gamma(5/2)+{\Gamma(7/2)\over x \beta m} +{\Gamma(9/2)\over 4(\beta m x)^2}\right].\nonumber\\
\end{eqnarray}
All these terms are finite when we take the limit $\mu\to m^-$, hence beyond a critical density for a given temperature, we will have Bose-Einstein condensation.
It is interesting to test $T\to 0^+$ limit as well. We see that as we let $\beta\to \infty$, the upper bound on the Bessel funtion implies that  none of the excited levels could be occupied.

Let us also remark on the two dimensional case,
following a similar analysis, we have
\begin{equation}
 \frac{m^{3/2}}{B\beta^{1/2}}\sum_{k=1}\frac{\sinh(k\beta \mu)}{k^{1/2}}K_{3/2}(k\beta m)   \leq n_{e}(T)\leq \frac{Am^{3/2}}{C\beta^{1/2}}\sum_{k=1}\frac{\sinh(k\beta \mu)}{k^{1/2}}K_{3/2}(k\beta m).
\end{equation}
Note that the Bessel function $K_{3/2}$ is given by
\begin{equation}
K_{3/2}(x)=\Big({\pi\over 2x}\Big)^{1/2}{ e^{-x}\over \Gamma(3/2)}[1+{1\over x}]>\Big({\pi\over 2x}\Big)^{1/2}{ e^{-x}\over \Gamma(3/2)}.
\end{equation}
This is a monotonically decreasing function hence, integral from $1$ to $\infty$ provides a lower bound;
\begin{equation}
  \frac{m^{3/2}}{B\beta}\int_1^\infty dx \sinh(x\beta \mu)\frac{e^{-x\beta m}}{x}  \leq n_{e}(T).
\end{equation}
For finite $\beta$  this integral is ultraviolet divergent if we let $\mu\to m^-$. Hence there is no need for  condensation in two dimensions.

We note the alternative expression for the excited density:
\begin{eqnarray}
\frac{\partial}{\partial (\beta\mu)}
\sum\limits_{k=-\infty}^\infty \int_0^\infty\
\frac{ds}{s^{3/2}}\,\frac{1}{2\sqrt{\pi}}\, e^
{\frac{-k^{2}}{4s}+k\beta\mu}\,\left(\frac{1}{V}\,\textrm{Tr}'
e^{s\beta^{2}\Delta}\right) e^{-s\beta^{2}m^{2}}.\nonumber\\
\end{eqnarray}
Again, we note that the $k$ sum gives the trace of the heat kernel on
$S^{1}$ coupled to the vector potential $a=2s\beta\mu d\theta$. Thus
we get
\begin{eqnarray}
n_e(T)=\frac{\partial}{\partial
(\beta\mu)}\int_{0}^{\infty}\frac{ds}{s^{3/2}}\,e^{-s\beta^{2}(m^{2}-\mu^{2})}\,\left(\frac{1}{V}\textrm{Tr}'
e^{s\beta^{2}\Delta}\right)\left(\frac{\sqrt{\pi}}{2\pi}\,\textrm{Tr}'\,e^{\frac{1}{4s}\Delta_{S^{1}}(a)}\right).
\end{eqnarray}
Here
\begin{equation}
    \Delta_{S^{1}}(a)=-(-i\frac{d}{d\theta}-a)^{2},
\end{equation}
is the Laplacian on $S^{1}$ coupled to the vector potential $a$
($\theta$ is the coordinate on $S^{1}$).

On the other hand the $k$ sum can also be calculated using the
Jacobi theta function of the third kind
\begin{equation}
    \sum_{k=-\infty}^{\infty}e^{-\frac{k^{2}}{4t}+k\beta\mu}=\theta_{3}\left(\frac{\beta\mu}{2i},e^{-\frac{1}{4t}}\right).
\end{equation}
Thus we get the alternative expression
\begin{eqnarray}
n_e(T)=\frac{\partial}{\partial
(\beta\mu)}\int_{0}^{\infty}\frac{ds}{s^{3/2}}\,\frac{1}{2\sqrt{\pi}}\,e^{-s\beta^{2}m^{2}}\,\left(\frac{1}{V}\textrm{Tr}'
e^{s\beta^{2}\Delta}\right)\,\theta_{3}\left(\frac{\beta\mu}{2i},e^{-\frac{1}{4s}}\right).
\end{eqnarray}

\section{Bogoliubov Theory on a Compact Riemannian Manifold}

 The many-body Hamiltonian with a hard-core repulsive potential is
given by \cite{Bogoliubov}, \cite{HuangYang}, \cite{LeeYang},
\cite{LeeHuangYang}
\begin{equation}\label{Hamiltonian}
    H^{\prime}=\int d\mu_{g}
    \,[\phi^{\dag}(x)h\phi(x)+\frac{u_{0}}{2}\,\phi^{\dag}(x)\phi^{\dag}(x)\phi(x)\phi(x)]\;\;\;(u_{0}>0).
\end{equation}
It is convenient to include the chemical potential in the
Hamiltonian and define
\begin{equation}
    H=H^{\prime}-\mu N.
\end{equation}
We will study this Hamiltonian and the Bose-Einstein condensation by
applying the curved space version of the standard Bogoluibov theory
\cite{Bogoliubov} to it. In flat space the Bogoliubov approximation
consists of three steps. First, one replaces the zero energy (ground
state) creation and annihilation operators by their coherent state
lower symbols and then one expands the Hamiltonian around the
c-number background obtained in the first step, ignoring third and
higher order terms in the fluctuation. The resulting Hamiltonian is
quadratic in the creation and annihilation operators but not
diagonal. The final step is the diagonalization of this Hamiltonian
by the Bogoluibov transformation. The justification of the c-number
replacement was first given in \cite{Ginibre} and more recently (and
with less assumptions) in \cite{Lieb} (see also \cite{Suto1},
\cite{Suto2}). In Appendix B we will show that on any Riemannian manifold (without any restriction on the Ricci curvature), thanks to the Neumann boundary condition which implies the constancy of the ground
state wave-function, the c-number substitution is justified in a way
similar to the flat case discussed in \cite{Lieb}. On the other hand, in the semiclassical approximation of
the Hamiltonian the Neumann boundary condition and the constancy of
the wave-function again play important roles. Finally the
Bogoluibov transformation which is a purely algebraic manipulation
proceeds in the usual way.

Therefore our starting point will be the expansion of the field
operator $\phi(x)$ around the background
\begin{equation}
    \phi_{0}(x)=\sqrt{N_{0}}f_{0}=\sqrt{\frac{N_{0}}{V}}=\sqrt{n_{0}}
\end{equation}
as
\begin{equation}
    \phi(x)= \phi_{0}(x)+\eta(x),
\end{equation}
with
\begin{equation}
\eta(x)=\sum_{\sigma\neq 0}a_{\sigma}f_{\sigma}.
\end{equation}
Assuming the quantum fluctuations to be small we can approximate
$H^{\prime}$ as
\begin{eqnarray}
  H'_{eff} &=& \int d\mu_{g}\;\big(
    \phi_{0}(x)h\eta(x)+\eta^{\dag}(x)h_{0}\eta(x)
   +\frac{u_{0}}{2}\left[\phi_{0}^{4}+2\,\phi_{0}^{3}(\eta^{\dag}(x)+
   \eta(x))\right.\nonumber\\
   &&\left.\phi_{0}^{2}(\eta^{2}(x)+\eta^{\dag\,2}(x)+4\,\eta^{\dag}(x)\eta(x))
   \right]\big).
\end{eqnarray}
Expanding in terms of creation and annihilation operators and using
(\ref{ort1}) and (\ref{ort2}) we get
\begin{equation}
H'_{eff}=\frac{u_{0}n_{0}^{2}V}{2}+\sum_{\sigma\neq
0}[(\epsilon_{\sigma}+2u_{0}n_{0})a^{\dag}_{\sigma}a_{\sigma}+\frac{u_{0}n_{0}}{2}(a_{\sigma}^{2}+a_{\sigma}^{\dag\,2})].\label{nonren}
\end{equation}
Similarly, the number operator is given by
\begin{eqnarray}
  N &=& \int\,d\mu_{g}(x) \,\phi^{\dag}(x)\phi(x)\nonumber\\
  &=&\int\,d\mu_{g}(x)\, [\phi_{0}^{2}+\phi_{0}(\eta^{\dag}(x)+\eta(x))+\eta^{\dag}(x)\eta(x)]\nonumber\\
   &=&n_{0}V+\sum_{\sigma\neq 0}a^{\dag}_{\sigma}a_{\sigma}.
\end{eqnarray}
So,
\begin{equation}
    H'_{eff}-\mu N=\frac{u_{0}n_{0}^{2}V}{2}-\mu n_{0}V +\sum_{\sigma\neq
0}[(\epsilon_{\sigma}+2u_{0}n_{0}-\mu)a^{\dag}_{\sigma}a_{\sigma}+\frac{u_{0}n_{0}}{2}(a_{\sigma}^{2}+a_{\sigma}^{\dag\,2})].
\end{equation}
For a fixed $n_{0}$ the thermodynamic pressure
\begin{equation}
    \frac{1}{V}\ln \textrm{Tr}^{\prime} e^{-\beta(H'_{eff}-\mu N)}
\end{equation}
is maximized at the zeroth order in $\eta$ by $\mu=n_{0}u_{0}$. Here
$\textrm{Tr}^{\prime}$ means that the trace is taken over the states
with no quanta in the $f_{0}$ mode. With this value of $\mu$ we get
\begin{equation}
    H'_{eff}-\mu N=-\frac{u_{0}n_{0}^{2}V}{2} +\sum_{\sigma\neq
0}[(\epsilon_{\sigma}+u_{0}n_{0})a^{\dag}_{\sigma}a_{\sigma}+\frac{u_{0}n_{0}}{2}(a_{\sigma}^{2}+a_{\sigma}^{\dag\,2})].
\end{equation}
Therefore the thermal averages can be calculated using the effective
Hamiltonian
\begin{equation}
    H_{eff}=\sum_{\sigma\neq
0}[(\epsilon_{\sigma}+u_{0}n_{0})a^{\dag}_{\sigma}a_{\sigma}+\frac{u_{0}n_{0}}{2}(a_{\sigma}^{2}+a_{\sigma}^{\dag\,2})],
\end{equation}
which can be diagonalized by the Bogoliubov transformation
\begin{eqnarray}
    a_{\sigma}=(\sinh\xi_{\sigma})\; b^{\dag}_{\sigma}+(\cosh\xi_{\sigma})\;
    b_{\sigma},
\end{eqnarray}
where
\begin{eqnarray}
  \lambda_{\sigma}&:=&\epsilon_{\sigma}+u_{0}n_{0} = \omega_{\sigma}\cosh 2\xi_{\sigma}, \\
  u_{0}n_{0} &=& -\omega_{\sigma}\sinh 2\xi_{\sigma},\\
  \omega_{\sigma} &:=&
  \sqrt{(\epsilon_{\sigma}+u_{0}n_{0})^{2}-(u_{0}n_{0})^{2}}.
\end{eqnarray}
The last equation is the curved space analog of the Bogoliubov
dispersion relation. The resulting diagonal Hamiltonian is
\begin{eqnarray}
  H_{eff} &=&\sum_{\sigma\neq
0}\left[\lambda_{\sigma}\cosh 2\xi_{\sigma}+u_{0}n_{0}\sinh 2\xi_{\sigma}\right]b_{\sigma}^{\dag}b_{\sigma}+E_{0}\nonumber  \\
   &=&\sum_{\sigma\neq 0}
   \omega_{\sigma}b_{\sigma}^{\dag}b_{\sigma}+E_{0},
\end{eqnarray}
where $E_{0}$ is the ground state energy
\begin{eqnarray}
    E_{0}&=&-\frac{1}{2}\sum_{\sigma=1}^{\infty}(\lambda_{\sigma}-\omega_{\sigma})\nonumber\\
    &=&-\frac{1}{2}\sum_{\sigma=1}^{\infty}\frac{(u_{0}n_{0})^{2}}{2}
    \frac{1}{\epsilon_{\sigma}}+O(\frac{1}{\epsilon_{\sigma}^{2}}).
\end{eqnarray}
Now using the lower eigenvalue bound (\ref{liyau}) we see that for
$k\geq 2$ and $d=3,2$
\begin{equation}
\sum_{\sigma=1}^{\infty}\frac{1}{\epsilon_{\sigma}^{k}}
\end{equation}
is convergent. On the other hand, using (\ref{colbois}) we see that
the $k=1$ sum is divergent for $d=3,2$.

Thus we arrive at the renormalized ground state energy.
\begin{equation}
E_{0}=-\frac{1}{2}\sum_{\sigma=1}^{\infty}\left[(\epsilon_{\sigma}+u_{0}n_{0})-\sqrt{\epsilon_{\sigma}^{2}+2u_{0}n_{0}\epsilon_{\sigma}}-
\frac{(u_{0}n_{0})^{2}}{2\epsilon_{\sigma}}\right].
\end{equation}
Indeed, a closer inspection of the  original Hamiltonian given by equation (\ref{nonren}) reveals that the bare coupling constant  should be replaced by a renormalized coupling constant. The bare coupling can be solved  order by order in
terms of the renormalized coupling constant $u_0^r$, to make the effective potential well-defined, and only the first term of (\ref{nonren}), within perturbation theory, should be modified by this subtraction. This gives us the desired term.  From now on, not to complicate matters we will interpret $u_0$ as  the renormalized coupling constant.

Finally, the ground state is the coherent state (see e.g. \cite{Umezawa})
\begin{equation}
    |\,\Omega\,\rangle=\mathcal{N}\prod_{\sigma=1}^{\infty}e^{-\frac{1}{2}\tanh
    \,\xi_{\sigma}a_{\sigma}^{\dag\,2}}|0\rangle
\end{equation}
with the normalization constant
\begin{equation}
\mathcal{N}=\prod_{\sigma=1}^{\infty}\frac{1}{\sqrt{\cosh\xi_{\sigma}}}=\left[\prod_{\sigma=1}^{\infty}1-\frac{\lambda_{\sigma}-\omega_{\sigma}}
{\lambda_{\sigma}+\omega_{\sigma}}\right]^{1/4}.
\end{equation}
Since $0\leq
(\lambda_{\sigma}-\omega_{\sigma})(\lambda_{\sigma}+\omega_{\sigma})^{-1}<1$
the convergence of the product is equivalent to the convergence of
the series
\begin{equation}
    \sum_{\sigma=1}^{\infty}\frac{\lambda_{\sigma}-\omega_{\sigma}}
{\lambda_{\sigma}+\omega_{\sigma}}.
\end{equation}
But,
\begin{eqnarray}
  \sum_{\sigma=1}^{\infty}\frac{\lambda_{\sigma}-\omega_{\sigma}}
{\lambda_{\sigma}+\omega_{\sigma}} &\leq&
\sum_{\sigma=1}^{\infty}\frac{\lambda_{\sigma}-\omega_{\sigma}}
{\epsilon_{\sigma}}=\sum_{\sigma=1}^{\infty}\frac{\lambda_{\sigma}-\omega_{\sigma}-\frac{(u_{0}n_{0})^{2}}{2\epsilon_{\sigma}}}
{\epsilon_{\sigma}}+ \sum_{\sigma=1}^{\infty}\frac{(u_{0}n_{0})^{2}}{2\epsilon_{\sigma}^{2}}.\nonumber\\
\end{eqnarray}
By the eigenvalue estimates given above, the last series is
convergent. On the other hand for $\sigma$ large enough
\begin{equation}
    \frac{\lambda_{\sigma}-\omega_{\sigma}-\frac{(u_{0}n_{0})^{2}}{2\epsilon_{\sigma}}}
{\epsilon_{\sigma}}\leq
\lambda_{\sigma}-\omega_{\sigma}-\frac{(u_{0}n_{0})^{2}}{2\epsilon_{\sigma}}.
\end{equation}
Combining this with our discussion of the ground state energy we see
that the first series is convergent as well. Thus we conclude
$\mathcal{N}<\infty$.

\section{Ground State Energy and Finite Size Effects}

To understand the ground state energy for three dimensions  better, 
 we will now express it in terms of the heat kernel. Consider the first two terms before the renormalization
--there is  $-1/2$ in front of the whole expression.   Let's write $s$ for $\epsilon_\sigma$ and $a$ for $u_0n_0$, for simplicity.
When we get $e^{-st}$  the sum over $\sigma$ gives us $\textrm{Tr}\,e^{\Delta t}$. Consider the expression
\begin{equation}
 s+a-\sqrt{s^2+2sa}.
\end{equation}
This function is equal to the shifted version $s\to s+a$
of
\begin{equation}
 s-\sqrt{s^2-a^2},
\end{equation}
which is equal to
\begin{equation}
 s-\sqrt{s^2-a^2}={a^2\over s+\sqrt{s^2-a^2}}
\end{equation}
Now we note the following integral representation ( see pg 326 line 8 of \cite{GR}):
\begin{equation}
\int_0^\infty {x dx\over [x^2+2sx+s^2-a^2]^{3/2}}=
{1\over s+\sqrt{s^2-a^2}}.
\end{equation}
This function is equal to a Laplace transform (for all Laplace transforms used in this paper see \cite{Abr})
\begin{eqnarray}
{1\over [x^2+2sx+s^2-a^2]^{3/2}}&=&{1\over [(x+s)^2-a^2]^{3/2}}\nonumber\\
&=&\int_0^\infty dt e^{-(s+x)t}
{\sqrt{\pi}\over \Gamma(3/2)}\Big({t\over 2a}\Big)I_1(at).
\end{eqnarray}
We simplify the numerical parts and again shift $s$ to $s+a$ to finally write
\begin{equation}
\int_0^\infty dt\int_0^\infty x dx  e^{-(s+a+x)t}
\Big({t\over a}\Big)I_1(at)=\int_0^\infty {dt\over at} I_1(at) e^{-at} e^{-st}.
\end{equation}
In the original sums we recognize now $\sum_\sigma e^{-\epsilon_\sigma t}$ as integral of the heat kernel,  which comes from the Laplace transform variable,
\begin{equation}
\int_0^\infty dt\int_0^\infty x dx  e^{-(s+a+x)t}
\Big({t\over a}\Big)I_1(at)=\int_0^\infty {dt\over at} I_1(at) e^{-at} \textrm{Tr}\,e^{\Delta t}.
\end{equation}
Let us see the convergence properties, since $\textrm{Tr}\,e^{\Delta t}=O(t^{3/2})$, and $I_1(t)\sim e^t/\sqrt{t}$,  there is no divergence as $t\to \infty$. However, as $t\to 0^+$ which corresponds to the ultraviolet properties,
we have $I_1(t)\sim t/2$ we get
\begin{equation}
   {1\over at} I_1(at) e^{-at} \textrm{Tr}\,e^{\Delta t} \sim {1\over at} {at\over 2} {1\over t^{3/2}}\ \ {\rm as} \ \ \ t\to 0^+ .
\end{equation}
which diverges,
yet if we subtract from this expression $1/2$
we have
\begin{equation}
   \Big({1\over at} I_1(at) e^{-at}-{1\over 2}\Big) \textrm{Tr}\,e^{\Delta t} \sim {1\over at} {at\over 2}(at) {1\over t^{3/2}}\ \ {\rm as} \ \ \ t\to 0^+ .
\end{equation}
which becomes convergent.
The subtracted term, with the overall $a^2$ term being inserted back again, is indeed
\begin{equation}
-{a^2\over 2}\textrm{Tr}\,e^{\Delta t}=-{a^2\over 2}\int_0^\infty dt \sum_\sigma e^{-\epsilon_\sigma t}=-\sum_\sigma {a^2\over 2\epsilon_\sigma}.
\end{equation}
Note that the subtraction does not lead to an infrared divergence, i.e. an ultraviolet divergence in the $t$ variable,  thanks to $t^{-3/2}$ behavior of the heat kernel.

Rewriting the ground state energy by moving the chemical potential part to the right hand side we get
\begin{equation}
E_g={u_0n_0^2 V\over 2}-{a^2 V\over 2}\int_0^\infty dt  \Big({1\over at} I_1(at) e^{-at}-{1\over 2}\Big) {1\over V}\textrm{Tr}\,e^{\Delta t}
\end{equation}
here $a=u_0n_0$. Note that we can scale $a$ in the integrals to get
\begin{equation}
E_g={u_0n_0^2 V\over 2}-{a V\over 2}\int_0^\infty dt  \Big({1\over t} I_1(t) e^{-t}-{1\over 2}\Big) {1\over V}\textrm{Tr}\,e^{\Delta t/a}
\end{equation}
Let us recall the formula 3.534 from \cite{GR};
\begin{equation}
{I_1(t)\over t}={ 2\over \pi} \int_0^1 dx \sqrt{1-x^2} \cosh(tx)
\end{equation}
moreover we have
\begin{equation}
\int_0^1 dx \sqrt{1-x^2} = {\pi\over 4}
\end{equation}
As a result we rewrite the ground state energy as
\begin{equation}
E_g={u_0n_0^2 V\over 2}-{a V\over  \pi}\int_0^\infty dt  \int_0^1 dx \sqrt{1-x^2} \Big(\cosh(tx)e^{-t}-1 \Big) {1\over V}\textrm{Tr}\,e^{\Delta t/a}
\end{equation}
or reorganizing this as
\begin{equation}
E_g={u_0n_0^2 V\over 2}+{a V \over 2\pi}\int_0^\infty dt  \int_0^1 dx\, F(t,x)  {1\over V}\textrm{Tr}\,e^{\Delta t/a}
\end{equation}
where
\begin{equation}
  F(t,x)=\sqrt{1-x^2}\Big(\underbrace{1-e^{-t(1-x)}}_{>0}+\underbrace{1-e^{-t(1+x)}}_{>0} \Big)>0 \;\;\textrm{for}\;\;0\leq x\leq 1.
\end{equation}
This shows that the part multiplying the heat kernel is strictly  positive, hence we may apply upper and lower bounds for the heat kernel to find estimates.
Nevertheless we have {\it a new version of} the formula derived by Lee and Yang \cite{LeeYang} in the flat space, extended now to a manifold,  for the ground state energy.

How can we test this in the flat space? Let us write down the heat kernel in the flat space into the formula assuming $V\to\infty$ for simplicity and {\it putting back again the physical values for the energy eigenvalues}, i.e. heat kernel corrected  with a factor of $\hbar^2/2m$:
\begin{equation}
E_g={u_0n_0^2 V\over 2}+{a V \over 2\pi }\int_0^\infty dt  \int_0^1 dx\, F(t,x) {a^{3/2}2^{3/2} m^{3/2}\over \hbar^3(4\pi t)^{3/2}}
\end{equation}
We now recall that
\begin{equation}
\int_0^\infty {dt\over t^{3/2}} \left( 1-e^{-t(1\pm x)}\right)=2 \sqrt{\pi}(1\pm x)^{1/2}.
\end{equation}
Thus we get
\begin{eqnarray}
E_g&&={u_0n_0^2 V\over 2}\left[1+{ 2^{5/2}(\pi)^{1/2}\over  \pi n_0} {a^{3/2} m^{3/2}\over \hbar^3 (4\pi )^{3/2}}  \int_0^1 dx \sqrt{1-x^2}\right.\nonumber\\
&&\;\;\;\;\;\;\;\;\;\;\;\;\;\;\;\left. \left((1+x)^{1/2}+(1-x)^{1/2}\right)\right]
\end{eqnarray}
The integral can be written as
\begin{equation}
\int_0^1 dx\Big( (1-x)^{1/2} (1+x)+(1+x)^{1/2} (1-x)\Big)
\end{equation}
by the formula 3.214 in \cite{GR} and is equal to $2^{3/2+2-1} B(3/2, 2)$ ($B$ is the beta function).
\begin{equation}
E_g={u_0 n_0 N\over 2}\left[1+{2^{5/2}\over \pi^{1/2}n_0}  \left( 2^{5/2}B(3/2,2)\right) {a^{3/2} m^{3/2}\over \hbar^3(4\pi )^{3/2}}\right]
\end{equation}
Now note that $u_0=4\pi \sigma \hbar^2/m$. Hence in terms of the scattering cross section,
 the energy per particle becomes,
\begin{equation}
{E_g\over N}={2\pi \hbar^{2}\sigma n_0 \over m}\left[1+{2^{5/2}\over \pi^{1/2} }  \left( 2^{5/2}B(3/2,2)\right) {n_0^{1/2} \sigma^{3/2}}\right].
\end{equation}
We have $B(3/2, 2)=4/15$, thus
\begin{equation}
{E_g\over N}={2\pi \hbar^{2}\sigma n_0 \over m}\left[1+{128\over 15 \pi^{1/2} } {(n_0 \sigma^3)^{1/2}}\right].
\end{equation}
the well-known result of Lee and Yang is recovered.

The expression we have found opens up two different directions: first is that we may find upper and lower bound estimates for the ground state energy using the known bounds of the heat kernel.  We will now present an upper bound in the thermodynamic limit for the ground state energy of the interacting system.
To this effect we use formula (11) for the upper bound of the trace of the heat kernel, place this into the expression for the ground state energy, replace the upper limit of $t$-integral  by infinity  and estimate the remainder term; this results in
\begin{equation}
E_g\leq C(3) E^{flat}_g+C(3){a^2 \over 2\pi}\int_{D_M^2}^\infty dt  \int_0^1 dx\, F_a(t,x)[{D_M^3\over t^{3/2}}+1]  
\end{equation}
where we use the natural units $\hbar=1, 2m=1$ again for simplicity, and $C(3)$ is a constant which depends on the dimension $d=3$ only, $D_M$ is the diameter of the box as discussed before. We have
\begin{equation}
F_a(t,x)=\sqrt{1-x^2} [ 1- e^{-at(x-1)}+1-e^{-at(x+1)}]
\end{equation}
Note that we have 
\begin{equation}
F_a(t,x) <2\sqrt{1-x^2} [ 1- e^{-2at}]
\end{equation}
 We now argue that the last term is a lower order correction,  as we take the limit $V\to \infty$. 
It is easy to see that the constant term is not of much importance, so we focus on the $1/t^{3/2}$ term. This term can be written as,
\begin{eqnarray}
C(3){a^2  D_M^3 \over \pi}\int_{D_M^2}^\infty dt  \int_0^1 &dx& (1-x^2)^{1/2} \int_0^{2a} d\eta e^{-\eta t}t{1\over t^{3/2}}\nonumber\cr 
&=&C(3){a^2  D_M^3 }\int_0^{2a}d\eta  \int_{D_M}^\infty e^{-\eta u^2}  du\nonumber
\end{eqnarray}
We now observe that 
\begin{equation}
\int_0^{2a}d\eta  \int_{D_M}^\infty e^{-\eta u^2}  du<\int_0^{2a}d\eta  e^{-\eta D_M^2} \int_0^\infty e^{-\eta u^2} du=\int_0^{2a}d\eta  {e^{-\eta D_M^2}\over \sqrt{\eta}}\nonumber
\end{equation}
As a result we see that this term is less than
\begin{equation}
C(3){a^2  D_M^3}\int_0^{2a}d\eta  {e^{-\eta D_M^2}\over \sqrt{\eta}}< C(3){a^2  D_M^3\sqrt{\pi}\over 2}{1\over D_M}\nonumber
\end{equation}
which goes to zero for $E_g/N$ when we take the limit $V\to \infty$, hence it is a  lower order correction when we assume $V\approx O(D_M^3)$.

As a second alternative, we may write down the finite volume version of the heat kernel, and find the corrections which may be coming from the boundary effects.
A first guess would be to apply an asymptotic expansion for the short time behaviour of the heat kernel.  Finite size effects for the condensation of the free gas from this perspective  is presented in \cite{Tomsfinite}, following the idea in this work we will obtain finite size effects of the weakly interacting condensate.
Let us digress a bit on the asymptotics of the Euclidean Domains which denote as  $\Omega$.
In three dimensions the precise asymptotics for the Neumann problem is given by \cite{Gilkey}
\begin{equation}
{\rm Tr}\, e^{\Delta t}={V(\Omega)\over (4\pi t)^{3/2}}+ {A(\Omega)\over 16\pi t} +{1\over 12\pi^{3/2}t^{1/2}}\int_{\partial \Omega} H(\omega) d\omega+...\quad {\rm as}\  t\to 0^+
\end{equation}
 where $A(\Omega)$ and $H(\omega)$ refer to the area and mean curvature of the surface respectively.
To use this in our expression we rewrite it as,
\begin{eqnarray}
E_g&\approx&{u_0n_0^2 V(\Omega)\over 2}+{a V(\Omega) \over 2\pi}\int_0^\infty dt  \int_0^1 dx \left[ F(t,x){ a^{3/2}\over (4\pi t)^{3/2}}\right]\nonumber\\
 &&+{a \over 2\pi}\int_0^1 dt  \int_0^1 dx F(t,x) \Big({A(\Omega)a\over 16\pi t} +{a^{1/2}\over 12\pi^{3/2}t^{1/2}}\int_{\partial \Omega} H(\omega) d\omega+...\Big)\nonumber\\
&\ & \ \ +{a V(\Omega) \over 2\pi}\int_1^\infty dt  \int_0^1 dx\,F(t,x)\Big( {1\over V(\Omega)}{\rm Tr}\, e^{\Delta t/a}-{a^{3/2}\over (4\pi t)^{3/2}}\Big)\nonumber
\end{eqnarray}
Now we claim that the last term of this expression can be made smaller than the others, hence of less significance. We leave this question to a subsequent publication. Retaining the terms in the above expansion except the last part, we find finite size corrections to the interacting Bose-Einstein condensate.

Finally we can also calculate the chemical potential, by taking the derivative with respect to the particle number.
This gives us,
\begin{eqnarray}
&&{\partial E_g\over\partial  N}={u_0n_0}+{u_0 n_0^2 \pi}\int_0^\infty dt  \int_0^1 dx \sqrt{1-x^2} F(t,x) {1\over V}\textrm{Tr}\,e^{\Delta t} \nonumber\\
&&+u_0^2n_0 \int_0^\infty dt \int_0^1 dx\, a\sqrt{1-x^2} \Big((1-x)e^{-at(1-x)}+(1+x)e^{-at(1+x)}\Big){1\over V}\textrm{Tr}\,e^{\Delta t}.\nonumber\\
\end{eqnarray}

\section{Heat Kernel Analysis of the Depletion of the Condensate}
We will now study the depletion coefficient in $3d$ and $2d$, first at zero then at finite temperature.
The number density of excited states
\begin{eqnarray}
  n_{e} = \frac{1}{V}\sum_{\sigma\neq 0}\langle a_{\sigma}^{\dag}a_{\sigma}\rangle
\end{eqnarray}
is expressed in terms of quasi-particle states as
\begin{eqnarray}
  n_{e} &=&\frac{1}{V} \sum_{\sigma\neq 0}\left[\sinh^{2}\xi_{\sigma}+\cosh\,2\xi_{\sigma}\langle b_{\sigma}^{\dag}b_{\sigma}\rangle+
  \frac{1}{2}\sinh 2\xi_{\sigma}\langle b_{\sigma}^{\dag\,2}+b_{\sigma}^{2}\rangle\right]\nonumber\\
   &=&\frac{1}{V}\sum_{\sigma\neq 0}\left[\cosh\,2\xi_{\sigma}\left(\frac{1}{2}+\langle
   b_{\sigma}^{\dag}b_{\sigma}\rangle\right)-\frac{1}{2}\right]\nonumber\\
&=&\frac{1}{V}\sum_{\sigma\neq 0}\left[
\frac{1}{2}\cosh\,2\xi_{\sigma}\coth\,\frac{\beta\omega_{\sigma}}{2}-\frac{1}{2}\right]\nonumber\\
&=&\frac{1}{2V}\sum_{\sigma\neq
0}\left[\frac{\lambda_{\sigma}}{\omega_{\sigma}}\,\coth\frac{\beta\omega_{\sigma}}{2}-1\right].
\end{eqnarray}
The zero temperature limit is
\begin{equation}
n_{e}=\frac{1}{2V}\sum_{\sigma\neq
0}\left[\frac{\lambda_{\sigma}}{\sqrt{\lambda_{\sigma}^{2}-(u_{0}n_{0})^{2}}}-1\right].
\end{equation}
Now noting the Laplace transform
\begin{equation}
    \frac{1}{\sqrt{\lambda_{\sigma}^{2}-(u_{0}n_{0})^{2}}}=\int_{0}^{\infty}dt\,e^{-\lambda_{\sigma}t}I_{0}(u_{0}n_{0}t),
\end{equation}
where $I_{\nu}$ is the modified Bessel function of order $\nu$, we
get
\begin{eqnarray}
\frac{\lambda_{\sigma}}{\sqrt{\lambda_{\sigma}^{2}-(u_{0}n_{0})^{2}}}&=&\int_{0}^{\infty}dt\,\left[-\frac{d}{dt}\,e^{-\lambda_{\sigma}t}\right]
I_{0}(u_{0}n_{0}t)\nonumber\\
&=&
1+u_{0}n_{0}\int_{0}^{\infty}dt\,e^{-\lambda_{\sigma}t}I_{1}(u_{0}n_{0}t).
\end{eqnarray}
Thus
\begin{eqnarray}\label{NE}
  n_{e} &=& \frac{u_{0}n_{0}}{2V}\int_{0}^{\infty}dt\,\left[\sum_{\sigma\neq
0}e^{-\epsilon_{\sigma}t}\right]\,e^{-u_{0}n_{0}t}\,I_{1}(u_{0}n_{0}t)\nonumber \\
&=&\frac{u_{0}n_{0}}{2}\int_{0}^{\infty}dt\,
\frac{1}{V}\,\textrm{Tr}'e^{-ht}\,e^{-u_{0}n_{0}t}\,I_{1}(u_{0}n_{0}t).
\end{eqnarray}

\textit{Finite Volume:} At finite $V$, we can use either (\ref{in})
or the simpler observation
\begin{equation}
\textrm{Tr}'e^{-ht}=\sum_{\sigma=1}^{\infty}e^{-t\epsilon_{\sigma}}\sim
e^{-t\epsilon_{1}},
\end{equation}
together with
\begin{equation}
e^{-u_{0}n_{0}t}\,I_{1}(u_{0}n_{0}t)\sim \frac{1}{\sqrt{2\pi
u_{0}n_{0}t}}\;\;\;\textrm{as}\;\;\;t\rightarrow\infty,
\end{equation}
and conclude that the upper limit of (\ref{NE}) is finite. As for
the lower limit of integration, we combine the short time asymptotic
of the heat kernel
\begin{equation}
\textrm{Tr}^{\prime}\,e^{-ht}\sim -1+ \frac{V}{(4\pi
t)^{d/2}},\;\;\;\textrm{as}\;\;\;t\rightarrow 0
\end{equation}
with
\begin{equation}
e^{-u_{0}n_{0}t}\,I_{1}(u_{0}n_{0}t)\sim
\frac{u_{0}n_{0}t}{2}\;\;\;\textrm{as}\;\;\;t\rightarrow 0,
\end{equation}
to conclude that the integral is convergent for $d=3$ and $d=2$.

\textit{Thermodynamic Limit:}
Using (\ref{important}) in the limit
$V,D_{M}\rightarrow\infty$, $D_{M}^{d}/V\rightarrow A$ we get
\begin{equation}
n_{e}\leq
\frac{u_{0}n_{0}}{2}\frac{A}{\,C^{d/2}}\,\Gamma\left(\frac{d}{2}+1\right)\int_{0}^{\infty}dt\,\frac{1}{t^{d/2}}
e^{-u_{0}n_{0}t}I_{1}(u_{0}n_{0}t).
\end{equation}
The integral is convergent for $d=3$.

On the other hand
\begin{equation}
n_{e}\geq
\frac{u_{0}n_{0}}{2}\frac{1}{B^{d/2}}\frac{d}{2}\int_{0}^{\infty}dt\,\Gamma\left(\frac{d}{2},\frac{t\,B}{V^{2/d}}\right)\frac{1}{t^{d/2}}\,
e^{-u_{0}n_{0}t}I_{1}(u_{0}n_{0}t).
\end{equation}
Again the integral is convergent for $d=3$ and $d=2$. Moreover, the integrand
is a positive, increasing function of $V$ and by the monotone
convergence theorem the limit $V\rightarrow\infty$ gives
\begin{equation}
n_{e}\geq
\frac{u_{0}n_{0}}{2}\frac{1}{B^{d/2}}\,\Gamma\left(\frac{d}{2}+1\right)\int_{0}^{\infty}dt\,\frac{1}{t^{d/2}}\,
e^{-u_{0}n_{0}t}I_{1}(u_{0}n_{0}t).
\end{equation}
Finally, upon the change of variable $s=u_{0}n_{0}t$ we see that the
bounds scale as $(u_{0}n_{0})^{d/2}$
\begin{equation}
   (u_{0}n_{0})^{d/2} m\gamma_{d}\leq
    n_{e}\leq (u_{0}n_{0})^{d/2} M\gamma_{d},
\end{equation}
where
\begin{equation}
    m=\frac{1}{2}\,
    \frac{1}{B^{d/2}}\,\Gamma\left(\frac{d}{2}+1\right),\,\,\,\,\,M=\frac{1}{2}\,
    \frac{A}{C^{d/2}}\,\Gamma\left(\frac{d}{2}+1\right),
\end{equation}
and
\begin{equation}
    \gamma_{d}=\int_{0}^{\infty}\frac{ds}{s^{d/2}}\,e^{-s}\,I_{1}(s).
\end{equation}
Thus we get, as in the flat case,
\begin{equation}
    \frac{n_{e}}{n_{0}}=O(u_{0}^{d/2}n_{0}^{d/2-1}).
\end{equation}
The smallness of the parameter $u_{0}^{d/2}n_{0}^{d/2-1}$ can now be
used as a criterion for the validity of the Gross-Pitaevskii
equation (see e.g. \cite{Pethick}).

\section{Depletion of the Condensate at Finite Temperature}

In order to analyze the depletion of the condensate at finite
temperatures we expand $N_{e}$ in terms of exponentials
\begin{equation}
    N_{e}(T)=\frac{1}{2}\sum_{\sigma\neq
    0}\left[\frac{\lambda_{\sigma}}{\omega_{\sigma}}-1+2\sum_{k=1}^{\infty}\frac{\lambda_{\sigma}}{\omega_{\sigma}}\,e^{-k\beta
    \omega_{\sigma}}\right].
\end{equation}
Noticing the Laplace transform,
\begin{eqnarray}
   \frac{e^{-k\beta
    \omega_{\sigma}}}{\omega_{\sigma}}=\frac{e^{-k\beta
    \sqrt{\lambda_{\sigma}^{2}-(u_{0}n_{0})^{2}}}}{\sqrt{\lambda_{\sigma}^{2}-(u_{0}n_{0})^{2}}}=\int_{k\beta}^{\infty}dt\,
    e^{-\lambda_{\sigma}t}
    I_{0}(u_{0}n_{0}\sqrt{t^{2}-k^{2}\beta^{2}}),
\end{eqnarray}
we find
\begin{eqnarray}
 \frac{\lambda_{\sigma}\,e^{-k\beta
    \omega_{\sigma}}}{\omega_{\sigma}}&=& e^{-\lambda_{\sigma}k\beta}+u_{0}n_{0}\int_{k\beta}^{\infty}dt\,e^{-\lambda_{\sigma}t}\,
    \frac{t\,I_{1}(u_{0}n_{0}\sqrt{t^{2}-k^{2}\beta^{2}})}{\sqrt{t^{2}-k^{2}\beta^{2}}}\\
&=&e^{-\lambda_{\sigma}k\beta}+u_{0}n_{0}\int_{0}^{\infty}dt\,e^{-\lambda_{\sigma}\sqrt{t^{2}+k^{2}\beta^{2}}}\,
    \,I_{1}(u_{0}n_{0}t).
\end{eqnarray}
Thus, we arrive at
\begin{equation}
    n_{e}(T)=n_{e}(0)+\tilde{n}_{e}(T),
\end{equation}
where
\begin{eqnarray}
   \tilde{n}_{e}(T)&=& \frac{1}{V}\sum_{\sigma\neq 0}\sum_{k=1}^{\infty}
    \left[e^{-\lambda_{\sigma}k\beta}+
    u_{0}n_{0}\,\int_{0}^{\infty}dt\,e^{-\lambda_{\sigma}\sqrt{t^{2}+k^{2}\beta^{2}}}\,
    \,I_{1}(u_{0}n_{0}t)
    \right]\nonumber \\
  & =&\sum_{k=1}^{\infty}\left[(\frac{1}{V}\textrm{Tr}'e^{k\beta\,\Delta})e^{-k\beta
  u_{0}n_{0}}+\right.\nonumber\\
   &&\left.u_{0}n_{0}\int_{0}^{\infty}dt\,(\frac{1}{V}\textrm{Tr}'e^{\Delta\,\sqrt{t^{2}+k^{2}\beta^{2}}})e^{-u_{0}n_{0}\sqrt{t^{2}+k^{2}\beta^{2}}}\,I_{1}(u_{0}n_{0}t)
   \,
    \right].\nonumber\\
\end{eqnarray}
We will now analyze this expression in the thermodynamic limit  in $3d$.
Using the heat kernel upper bound (\ref{important}) in the limit
$V\rightarrow\infty$ we get
\begin{eqnarray}
\tilde{n}_{e}(T)&\leq& C_{1}\sum_{k=1}^{\infty}\left[{1\over
(k\beta)^{3/2}}e^{-k\beta
  u_{0}n_{0}}+\right.\nonumber\\
   &\ &\left.u_{0}n_{0}\int_{0}^{\infty}dt\,{1\over (t^{2}+k^{2}\beta^{2})^{3/4}}e^{-u_{0}n_{0}\sqrt{t^{2}+k^{2}\beta^{2}}}\,I_{1}(u_{0}n_{0}t)
   \,
    \right].
\end{eqnarray}
Here
\begin{equation}
    C_{1}=\frac{A}{C^{3/2}}\Gamma\left(\frac{5}{2}\right).
\end{equation}
We will estimate each term separately, the first expression is
bounded by
\begin{equation}
C_1\sum_{k=1}^{\infty}{1\over (k\beta)^{3/2}}+ C_1\int_0^\infty {dk\over (\beta k)^{3/2}}(1- e^{-k\beta  u_{0}n_{0}}),
\end{equation}
which is equal to
\begin{equation}
C_1\sum_{k=1}^{\infty}{1\over (k\beta)^{3/2}}+ C_1{(u_0n_0)^{1/2}\over \beta}.
\end{equation}
The second part is somewhat more subtle, we first  apply the
subordination identity for the exponent and find that the second
becomes,
\begin{equation}
C_2 u_{0}n_{0}\int_{0}^{\infty}dt\,\sum_{k=1}^{\infty}{1\over
(t^{2}+k^{2}\beta^{2})^{3/4}}\int_0^\infty {ds\over
s^{3/2}}e^{-{1\over
4s}-s(u_{0}n_{0})^2(t^{2}+k^{2}\beta^{2})}\,I_{1}(u_{0}n_{0}t).
\end{equation}
Next we estimate the summation, again the terms of the sum are monotonically decreasing as the summand increases, hence the integral gives an upper bound
which we estimate separately;
\begin{eqnarray}
&&\sum_{k=1}^{\infty} {1\over (t^{2}+k^{2}\beta^{2})^{3/4}}e^{-(u_{0}n_{0})^2sk^{2}\beta^{2}}<  \int_0^\infty {dk\over (t^{2}+k^{2}\beta^{2})^{3/4}}e^{-(u_{0}n_{0})^2sk^{2}\beta^{2}}\nonumber\\
&<& \Big[ \int_0^\infty {dk\over (t^{2}+k^{2}\beta^{2})^{3/2}}\Big]^{1/2}\Big[ \int_0^\infty dk e^{-2s(u_on_0)^2k^2\beta^2}\Big]^{1/2}\nonumber\\
&<&C_3 {1\over s^{1/4}(u_0n_0)^{1/2}\beta^{1/2}}{1\over \beta^{1/2}t}.
\end{eqnarray}
Note that in the first integral we scale the variable $k$  with
$t\beta^{-1}$ and the second integral by $(\sqrt{s}\beta
u_0n_0)^{-1}$. We may place this estimate now to find the upper
bound,
\begin{equation}
{C_4 (u_0n_0)^{1/2}\over \beta}\int_0^\infty {dt\over
t}\int_0^\infty {ds\over s^{1+3/4}}e^{-{1\over
4s}-s(u_{0}n_{0})^2t^{2}}\,I_{1}(u_{0}n_{0}t).
\end{equation}
We recognize the modified bessel function, to rewrite this expression as,
\begin{equation}
{C_5 (u_0n_0)^{1/2}\over \beta}\int_0^\infty {dt\over t}(u_0n_0
t)^{3/4}K_{3/4}(u_{0}n_{0}t)\,I_{1}(u_{0}n_{0}t) .\end{equation}
Note that in the integral $ (u_0n_0)$ completely scales out. The
integral is of the type given in Prudnikov et al. \cite{prudnikov}
formula 2.16.28.3.
\begin{eqnarray}
 \int_0^\infty dx\,  x^{\rho-1} K_\mu(x)I_\nu(x)= \frac{ 2^{\rho-1}\Gamma({1\over 2}(\rho+\nu+\mu))\Gamma({1\over 2}(\rho+\nu-\mu))\Gamma(1-\rho)}
{\Gamma(1+{1\over 2}(-\rho+\nu+\mu))\Gamma(1+{1\over 2}(-\rho+\nu -\mu))}.\nonumber
\end{eqnarray}
for $|\mu|-\nu<\rho<1$. Hence we find that
\begin{equation}
\tilde{n}_{e}(T)< C_1\sum_{k=1}^{\infty}{1\over (k\beta)^{3/2}}+ C_6
{(u_0n_0)^{1/2}\over \beta} ,\end{equation} the last piece of which
will go to zero as $u_0\to 0^+$ and moreover  the full expression
will go to zero as $\beta\to \infty$.

Next, we will show that the Bogoliubov approximation at finite
temperature is actually inconsistent in $2d$, in accordance with the
Hohenberg-Mermin-Wagner teorem \cite{Mermin},\cite{Hohenberg},
rigorously established in spin systems or interacting bosons in flat
spaces. To see this, we will use lower bounds on the heat kernel
(\ref{important}) in the limit $V\rightarrow\infty$. We will further
estimate sums of monotonically decreasing expressions from below  by
integrating them from $1$ to infinity. As a result we see that $\tilde{n}_{e}(T)$ is larger than
\begin{eqnarray}
{1\over \beta} \int_1^\infty {dk\over k}  e^{-(u_0n_0)k\beta} +u_0n_0\int_0^\infty dt \int_1^\infty {dk\over \sqrt{t^{2}+k^{2}\beta^{2}}}\,e^{-u_{0}n_{0}\sqrt{t^{2}+k^{2}\beta^{2}}}\,I_{1}(u_{0}n_{0}t).\nonumber\\
\end{eqnarray}
Note that we may shift  $k$ to $k+1$ and replace all $k^2+2k+1$'s by $3k^2+1$ since $k^2+2k+1<3k^2+1$ for $k\geq 0$.
After scaling $k$ to $ k\sqrt{t^2+\beta^2}/\sqrt{3}\beta$ we find that $\tilde{n}_{e}(T)$ is larger than
\begin{eqnarray}
 {1\over \beta} \int_1^\infty {dk\over k}  e^{-(u_0n_0)k\beta} +{u_0n_0\over\sqrt{3} \beta}\int_0^\infty dt \int_0^\infty {dk\over \sqrt{1+k^{2}}}e^{-u_{0}n_{0}\sqrt{t^2+\beta^2}\sqrt{1+k^{2}}}\,I_{1}(u_{0}n_{0}t).
\end{eqnarray}
 We notice that
\begin{equation}
\int_0^\infty {dk e^{-a\sqrt{1+k^2}}\over \sqrt{1+k^2}}= K_0(a),
\end{equation}
hence the lower bound becomes,
\begin{eqnarray}
{1\over \beta}  \int_1^\infty {dk\over k}  e^{-(u_0n_0)k\beta} +{u_0n_0\over\sqrt{3} \beta}\int_0^\infty dt K_0(u_0n_0\sqrt{ t^2+\beta^2})\,I_{1}(u_{0}n_{0}t) <\tilde{n}_{e}(T).
\end{eqnarray}
Nevertheless the integral of Bessel functions is ultraviolet divergent--which reflects the infrared behavior of the theory in the heat kernel approach-as a result of the asymptotics of the bessel functions,
\begin{equation}
 K_0(x)\sim {e^{-x}\over \sqrt{x}} \ \ \ {\rm and} \ \ \ I_1(x)\sim {e^x\over \sqrt{x}}\ \ \ {\rm as}\ \  x\to \infty.
\end{equation}
This contradiction forces $n_0=0$ to be the only consistent choice.

\section*{Acknowledgement}

This work is supported by Bo\u{g}azi\c{c}i University BAP Project No. 6942.  We would like to thank Dieter Van den Bleeken and Yusuf G\"{u}l for useful discussions.

\section*{Appendix A}

Here we will show how the upper bound (\ref{liyau}) for the
eigenvalues follows from the upper estimate for the trace of the
heat kernel given in (\ref{bound1}) \cite{LiYau}, \cite{Don}, and how one gets the long time
behavior (\ref{in}) from the self-reproducing property of the heat
kernel.

Starting from the simple observation
\begin{equation}
    (\sigma+1)e^{-\epsilon_{\sigma}t}\leq \textrm {Tr}e^{\Delta t}\leq
    \tilde{C}(d)g(t),
\end{equation}
and using the upper bound
\begin{equation}
\textrm{Tr}\,e^{\Delta t}\leq
    \tilde{C}(d)g(t),
\end{equation}
we get
\begin{equation}\label{in2}
 (\sigma+1)\leq e^{\epsilon_{\sigma}t}\tilde{C}(d)g(t)
\end{equation}
for all positive $t$. Minimizing the right hand side we get
\begin{equation}
    \epsilon_{\sigma}g(t_{0})+g^{\prime}(t_{0})=0.
\end{equation}
Since $g^{\prime}(t)=0$ for $\sqrt{t}>D_{M}$, we see that
$\sqrt{t_{0}}<D_{M}$. Then we get
\begin{equation}
    t_{0}=\frac{d}{2\epsilon_{\sigma}}.
\end{equation}
Plugging this into (\ref{in2}) we get the desired bound
\begin{equation}
    \epsilon_{\sigma}\geq \frac{C(d)}{D_{M}^{2}}\,(\sigma+1)^{2/d}\geq
    \frac{C(d)}{D_{M}^{2}}\,\sigma^{2/d}.
\end{equation}

Let $K_{t}(x,y)=\langle x|e^{\Delta\,t}|y\rangle$ be the heat kernel for the Neumann problem on $M$.
Clearly $K_{t}(x,y)$ is a self-reproducing kernel. It is convenient
to define
\begin{equation}
\bar{K}_{t}(x,y)=K_{t}(x,y)-\frac{1}{V}.
\end{equation}
Since $V^{-1/2}$ is the eigenfunction of the Laplacian with zero
eigenvalue we have
\begin{equation}
    \frac{1}{\sqrt{V}}=\int_{M}d_{g}\mu(y)\,K_{t}(x,y)\frac{1}{\sqrt{V}}.
\end{equation}
Using this it is easy to see that $\bar{K}_{t}$ is also
self-reproducing
\begin{equation}
    \bar{K}_{t_{1}+t_{2}}(x,y)=\int_{M}d_{g}\mu(z)\,\bar{K}_{t_{1}}(x,z)\,\bar{K}_{t_{2}}(z,y).
\end{equation}
Now note that \cite{Wang}
\begin{eqnarray}
  \frac{\partial}{\partial t}\bar{K}_{t}(x,x) &=&\frac{\partial}{\partial t}\int_{M}d_{g}\mu(z)\,\bar{K}^{2}_{t/2}(x,z)\nonumber  \\
   &=&-\int_{M}d_{g}\mu(z)\,\bar{K}_{t/2}(x,z)\,h\,\bar{K}_{t/2}(x,z)\nonumber  \\
   &\leq
   &-\epsilon_{1}\int_{M}d_{g}\mu(z)\,\bar{K}^{2}_{t/2}(x,z)=-\epsilon_{1}\bar{K}_{t}(x,x).
\end{eqnarray}
In the last line the variational inequality is used (see e.g.
\cite{Chavel}). Integrating this inequality one finds that for
$t\geq t_{0}$
\begin{equation}
\bar{K}_{t}(x,x)\leq \bar{K}_{t_{0}}(x,x)e^{-\epsilon_{1}(t-t_{0})}.
\end{equation}
Fixing the value of $t_{0}$ we see that the diagonal elements of
$\bar{K}_{t}$ decay exponentially in time. Integrating over $x$ we
get
\begin{equation}
    \textrm{Tr}^{\prime}\,e^{\Delta\,t} \leq
    (\textrm{Tr}^{\prime}\,e^{\Delta\,t_{0}})e^{-\nu_{1}(t-t_{0})}.
\end{equation}

\section*{Appendix B}

Let $\{|z\rangle\,z\in N\}$ be an over-complete set labeled by the
points of a manifold $N$ with measure $d\mu(z)$. The lower symbol
$A_{L}(z)$ of an operator $A$ is the expectation value
\begin{equation}
A_{L}(z)=\langle z|A|z\rangle.
\end{equation}
On the other hand, if there exist a function $A_{U}(z)$ on $N$ such
that
\begin{equation}
    A=\int d\mu(z)\,A_{U}(z)|z\rangle\langle z|,
\end{equation}
then $A_{U}(z)$ is called the upper symbol of $A$. Here the equality
is in the weak sense.

Let
\begin{equation}
    |z\rangle=e^{-\frac{|z^{2}|}{2}+za_{0}^{\dag}}|0\rangle,\;\;\;z\in
    \mathbf{C},
\end{equation}
be the standard coherent states for the annihilation operator
$a_{0}$. This is an over-complete set relative to the measure
\begin{equation}
    d\mu(z)=\frac{dzdz^{*}}{\pi}.
\end{equation}
The following list of the symbols of various combinations of
creation and annihilation operators is useful in calculating the
lower and upper symbols of the Hamiltonian.
\begin{eqnarray}
\begin{tabular}{|l|l|l|}
  \hline
   $A$& $A_{L}$ & $A_{U}$ \\ \hline
  $a_{0}$ & $z$ & $z$ \\
  $a_{0}^{\dag}$ & $z^{*}$ & $z^{*}$ \\
  $a_{0}^{2}$ & $z^{2}$ & $z^{2}$ \\
  $a_{0}^{\dag\,2}$ & $z^{*\,2}$ & $z^{*\,2}$ \\
  $a_{0}^{\dag}a_{0}$ & $|z|^{2}$ & $|z|^{2}-1$ \\
  $a_{0}^{\dag\,2}a_{0}^{2}$ & $|z|^{4}$ & $|z|^{4}-4|z|^{2}+2$ \\
  \hline
\end{tabular}
\end{eqnarray}
We assume our Hamiltonian includes the chemical potential
\begin{equation}
    H=H^{\prime}-\mu N,
\end{equation}
where $H^{\prime}$ is given by (\ref{Hamiltonian}). In the
Hamiltonian $H$ we replace every monomial of the form $a_{0}^{\dag
n}a_{0}^{m}$ first by its lower and then by its upper symbol and
thus obtain the two Hamiltonians $H_{L}(z,z^{*})$ and
$H_{U}(z,z^{*})$. Let $Z_{L}(\beta,\mu)$ and $Z_{U}(\beta,\mu)$ be
the corresponding grand canonical partition functions integrated
over $z$,
\begin{equation}
Z_{L,U}(\beta,\mu)=\int\frac{dzdz^{*}}{\pi}\,Z_{L,U}(\beta,\mu,z,z^{*}),
\end{equation}
where
\begin{equation}
    Z_{L,U}(\beta,\mu,z,z^{*})=\textrm{Tr}^{\prime} e^{-\beta
H_{L,U}(z,z^{*})}.
\end{equation}
Here $\textrm{Tr}^{\prime}$ means that the trace is taken over the
states with no excitations in the $f_{0}$ mode. Then we have the
following inequalities
\begin{equation}\label{JBL}
    Z_{L}(\beta,\mu)\leq Z(\beta,\mu)\leq Z_{U}(\beta,\mu).
\end{equation}
The first inequality is the Jensen's inequality and the second is
the Berezin-Lieb inequality \cite{Berezin1}, \cite{Berezin2},
 \cite{Simon}, \cite{Baumgartner}, \cite{Lieb0}, \cite{Lieb}. These inequalities
are valid on any manifold since geometry does not play any role
whatsoever in their derivations.

Comparing $H_{L}$ and $H_{U}$ we see that
\begin{eqnarray}\label{boundel}
   \delta&=& H_{U}(z,z^{*})-H_{L}(z,z^{*})=\mu+\frac{u_{0}}{4V}(-4|z|^{2}+2-4\sum_{\sigma\neq 0}a_{\sigma}^{\dag}a_{\sigma})\nonumber\\
   &=&\mu+\frac{u_{0}}{4V}(2-4N_{L}).
\end{eqnarray}
In deriving this we used the list of symbols given above and the
fact that $f_{0}=V^{-1/2}$.

Thus we find
\begin{eqnarray}
    \textrm{Tr}^{\prime}\,e^{-\beta H_{U}(z,z^{*})}&=&\textrm{Tr}^{\prime}\,e^{-\beta
    (H_{L}(z,z^{*})+\mu+\frac{u_{0}}{V}-\frac{2u_{0}}{V}N_{L})}\nonumber\\
    &=& e^{-\beta(\mu+\frac{u_{0}}{V})}
\textrm{Tr}^{\prime}\,e^{-\beta
(H_{L}(z,z^{*})-\frac{2u_{0}}{V}N_{L})},
\end{eqnarray}
or
\begin{equation}
    Z_{U}(\mu,\beta)=e^{-\beta(\mu+\frac{u_{0}}{V})}Z_{L}\left(\mu+\frac{2u_{0}}{V},\beta\right).
\end{equation}
So
\begin{eqnarray}
    \lim_{V\rightarrow \infty}\frac{1}{V}\ln Z_{U}(\mu,\beta)&=&\lim_{V\rightarrow \infty}\frac{1}{V}\left[-\beta(\mu+\frac{u_{0}}{V})\right]+\frac{1}{V}\ln Z_{L}\left(\mu+\frac{2u_{0}}{V},\beta\right)\nonumber\\
    &=&\lim_{V\rightarrow \infty}\frac{1}{V}\ln Z_{L}(\mu,\beta).
\end{eqnarray}
Then from (\ref{JBL}) we get the equality of the pressures in the thermodynamic limit
\begin{eqnarray}
    \frac{1}{V}\ln Z_{U}(\mu,\beta)=\frac{1}{V}\ln Z(\mu,\beta)
    =\frac{1}{V}\ln Z_{L}(\mu,\beta).
\end{eqnarray}

Let $z_{0}$ be the value of $z$ for which $Z_{L}(\beta,\mu,z,z^{*})$
is maximum. Then the integrals in the above expressions localize around $z_{0}$ and the following inequalities hold \cite{Lieb}
\begin{equation}
     \frac{1}{V}\ln
     Z_{L}(\mu,\beta,z_{0},z_{0}^{*})\leq \frac{1}{V}\ln Z(\mu,\beta)
     \leq  \frac{1}{V}\ln
     Z_{U}(\mu,\beta,z_{0},z_{0}^{*})+O\left(\frac{\ln
     V}{V}\right).
\end{equation}
Note that the usual choice $\mu=u_{0}n_{0}$ in the Bogoliubov theory is in accordance with this result. Again using (\ref{boundel}), in the thermodynamic limit we obtain
\begin{equation}
    \frac{1}{V}\ln Z(\mu,\beta)= \frac{1}{V}\ln
    Z_{L}(\mu,\beta,z_{0},z^{*}_{0}).
\end{equation}

\end{document}